# Theoretical and Practical Limits of Signal Strength Estimate Precision for Kolmogorov-Zurbenko Periodograms with Dynamic Smoothing

**Barry Loneck[*], Igor Zurbenko, and Edward Valachovic**

**Department of Epidemiology and Biostatistics**
**School of Public Health**
**The University at Albany**
**State University of New York**

## ABSTRACT

This investigation establishes the theoretical and practical limits of signal strength estimate precision for Kolmogorov-Zurbenko periodograms with dynamic smoothing and compares them to those of standard log-periodograms with static smoothing. Previous research has established the sensitivity, accuracy, resolution, and robustness of Kolmogorov-Zurbenko periodograms with dynamic smoothing in estimating *signal frequencies*. However, the precision with which they estimate *signal strength* has never been evaluated. To this point, the width of the confidence interval for a signal strength estimate can serve as a criterion for assessing the precision of such estimates – the narrower the confidence interval, the more precise the estimate. The statistical background for confidence intervals of periodograms is presented, followed by candidate functions to compute and plot them when using Kolmogorov-Zurbenko periodograms with dynamic smoothing. Given an identified signal frequency, a static smoothing window and its smoothing window width can be selected such that its confidence interval is narrower and, thus, its signal strength estimate more precise, than that of dynamic smoothing windows, all while maintaining a level of frequency resolution as good as or better than that of a dynamic smoothing window. These findings suggest the need for a two-step protocol in spectral analysis: computation of a Kolmogorov-Zurbenko periodogram with dynamic smoothing to detect, identify, and separate signal frequencies, followed by computation of a Kolmogorov-Zurbenko periodogram with static smoothing to precisely estimate signal strength and compute its confidence intervals.

[*] Corresponding Author Email: bloneck@albany.edu

**Theoretical and Practical Limits of Signal Strength Estimate Precision for Kolmogorov-Zurbenko Periodograms with Dynamic Smoothing**

This investigation establishes the theoretical and practical limits of signal strength estimate precision for Kolmogorov-Zurbenko periodograms with dynamic smoothing in contrast to standard log-periodograms with static smoothing. Spectral analysis is used to detect, identify, and separate signal frequencies as well as estimate signal strength in the periodic time series of a variable. Such analyses are typically done with periodograms with static smoothing windows that use a moving average with fixed window width across the frequency spectrum (Wei, 2006). However, Kolmogorov-Zurbenko periodograms (Zurbenko, 1986) with dynamic smoothing windows also use a moving average but with a window width that varies across the frequency spectrum, zooming in when a signal is present and zooming out when there is no signal. To this point, the DiRienzo-Zurbenko algorithm (DiRienzo, 1996; DiRienzo & Zurbenko, 1999) varies window width based on the proportion of total variance in the locale of a frequency, while the Neagu-Zurbenko algorithm (Neagu & Zurbenko, 2003; Zurbenko, 2001) varies window width based on the proportion of total departure from linearity in the locale of a frequency. Because of distinct advantages provided by a logarithmic transformation (e.g., comparisons independent of scaling units), Kolmogorov-Zurbenko periodograms with dynamic smoothing are, by default, log-periodograms. Previous work has established the sensitivity, accuracy, resolution, and robustness of Kolmogorov-Zurbenko periodograms with dynamic smoothing in estimating signal frequencies (Loneck et al., 2024; Zurbenko et al., 2020), but the theoretical and practical limits of their precision in estimating signal strength has not been established. However, estimates of both signal frequency and respective signal strength are required to frame a comprehensive model for a given periodic time series and, thereby, better predict future values of the periodic time series variable.

The width of the confidence interval for a signal strength estimate provides a means for assessing the precision of this estimate – the narrower the confidence interval, the more precise the estimate and the wider the confidence interval, the less precise the estimate. Confidence intervals for signal strength are also important because they can be used to determine whether there are statistically significant differences in signal strength for a single phenomenon in the same population at two different periods in time as well as discerning statistically significant differences in signal strength in the same phenomenon for two populations during the same period in time. Of course, if the analyst prefers, the p-value for a given frequency's observed signal strength vis-à-vis hypothesized strength can, likewise, be computed.

Utilizing an approach put forth by Priestley (1981) and by Koopmans (1995), then applied by DiRienzo (1996) and Wei (2006), the article begins with the statistical bases of confidence intervals for both static smoothing and dynamic smoothing. This is followed by descriptions of candidate functions for the *kza* package in the *R* platform such that confidence intervals for Kolmogorov-Zurbenko periodograms with dynamic smoothing can be computed and plotted; simulated time series datasets are used to demonstrate confidence intervals of Kolmogorov-Zurbenko periodograms with dynamic smoothing in the context of sensitivity and accuracy with a single signal, as well as in the context of resolution with a pair of signals that are close in frequency. Finally, the theoretical and practical limits for precision with which Kolmogorov-Zurbenko periodograms with dynamic smoothing and log-periodograms with static smoothing estimate signal strength are established and compared by contrasting the width of their respective confidence intervals; static smoothing windows included rectangular, Tukey-Hamming, Tukey-Hanning, Bartlett, and Parzen, while dynamic smoothing windows included DiRienzo-Zurbenko and Neagu-Zurbenko. The article concludes with a summary and with a recommendation for a two-step protocol for spectral analysis: computation of a Kolmogorov-Zurbenko periodogram with dynamic smoothing to detect, identify, and separate signal frequencies followed by computation of a Kolmogorov-Zurbenko periodogram with static smoothing to precisely estimate signal strength and compute confidence intervals. In this way, spectral analysts will be better able to model and predict future values of a given phenomenon with an underlying periodicity in time, in space, or in both time and space.

## 1. Confidence Intervals for Kolmogorov-Zurbenko Periodograms with Dynamic Smoothing

As noted in the Introduction, the precision with which a Kolmogorov-Zurbenko periodogram with dynamic smoothing estimates signal strength is reflected in the width of its confidence interval at a given frequency; the narrower the confidence interval, the more precise the estimate of signal strength. This section provides the statistical bases of confidence intervals for both static windows and dynamic windows, candidate functions to compute confidence intervals for dynamic windows, and demonstrations of confidence intervals for dynamic windows using simulated datasets.

### 1.1 Statistical Bases

In spectral analysis, the basis of confidence intervals for signal strength (i.e., spectral ordinates) has been provided by both Priestley (1981) and Koopmans (1995), with explication for dynamic smoothing by DiRienzo (1996) and explication for static smoothing by Wei (2006). In general, spectral ordinates across a spectrum are independent and identically distributed as:

$$\hat{f}(\omega_k) \sim f(\omega_k)\frac{\chi^2(2)}{2} \qquad (1.1.1)$$

where:

$\hat{f}(\omega_k)$ = spectral ordinate estimate at Fourier frequency $\omega_k$; and

$f(\omega_k)$ = spectral ordinate at Fourier frequency $\omega_k$

with $\chi^2$ having 2 degrees of freedom in the numerator and the denominator equal to the degrees of freedom, again 2.

When the sample spectrum is smoothed, the distribution of the smoothed spectral ordinate for the window $\mathcal{W}$ is:

$$\hat{f}_{\mathcal{W}}(\omega_k) \sim f(\omega_k)\frac{\chi^2(\upsilon)}{\upsilon} \qquad (1.1.2)$$

where:

$\hat{f}_{\mathcal{W}}(\omega_k)$ = smoothed spectral estimate at Fourier frequency $\omega_k$;
$f(\omega_k)$ = spectral ordinate at Fourier frequency $\omega_k$; and
$\upsilon$ = degrees of freedom for a given smoothing window.

Thus, Equation 1.1.2 implies that the $(1-\alpha)$ 100% confidence interval for a spectral ordinate is:

$$\frac{\nu\, \hat{f}_{\mathcal{W}}(\omega_k)}{\chi^2_{\alpha/2}(\nu)} \leq f(\omega) \leq \frac{\nu\, \hat{f}_{\mathcal{W}}(\omega_k)}{\chi^2_{(1-\alpha/2)}(\nu)} \qquad (1.1.3)$$

where $\alpha$ is the *upper* $\alpha$ proportion of the chi-square distribution. However, the width of this confidence interval varies with frequency. To eliminate the impact of frequency on confidence interval width, the analyst can use a logarithmic transformation of the spectrum estimate. Thus, we have

$$ln\left(\hat{f}_W(\omega_k)\right) + ln\left[\frac{\nu}{\chi^2_{\alpha/2}(\nu)}\right] \le ln(f(\omega_k)) \le ln(\hat{f}_W(\omega_k)) + ln\left[\frac{\nu}{\chi^2_{1-\alpha/2}(\nu)}\right] \quad (1.1.4)$$

As a result, the width of this confidence interval does not depend on frequency.

As a side note, the American Statistical Association now favors the reporting of p-values rather than hypothesis testing, so this confidence interval can be recast in order to report p-values when comparing the spectral density from one time period or population, $\hat{f}_{W_1}(\omega_k)$, to the spectral density for another time period or population, $\hat{f}_{W_2}(\omega_k)$. To determine the p-value with regard to whether $\hat{f}_{W_2}(\omega_k)$ is larger than $\hat{f}_{W_1}(\omega_k)$, we can use the upper bound of the confidence interval to determine whether

$$ln(\hat{f}_{W_1}(\omega_k)) + ln\left[\frac{\nu}{\chi^2_{1-\alpha/2}(\nu)}\right] \le ln(\hat{f}_{W_2}(\omega_k)) \quad (1.1.5)$$

By rearranging terms, we have

$$\nu\frac{\hat{f}_{W_1}(\omega_k)}{\hat{f}_{W_2}(\omega_k)} \le \chi^2_{1-\alpha/2}(\nu) \quad (1.1.6)$$

and, by implication, this leads to the p-value

$$Pr\left(\hat{f}_{W_2}(\omega_k) > \hat{f}_{W_1}(\omega_k)\right) = Pr\left(\chi^2(\upsilon) \le \nu\frac{\hat{f}_{W_1}(\omega_k)}{\hat{f}_{W_2}(\omega_k)}\right) \quad (1.1.7)$$

Similarly, when assessing whether $\hat{f}_{W_2}(\omega_k)$ is smaller than $\hat{f}_{W_1}(\omega_k)$, we can use the lower bound of the confidence interval to determine

$$Pr\left(\hat{f}_{W_2}(\omega_k) < \hat{f}_{W_1}(\omega_k)\right) = Pr\left(\chi^2(\upsilon) \ge \nu\frac{\hat{f}_{W_1}(\omega_k)}{\hat{f}_{W_2}(\omega_k)}\right) \quad (1.1.8)$$

Resuming discussion of confidence intervals, determination of the degrees of freedom for a static smoothing window is complex. In general, a given static smoothing window is the Fourier transform of its respective autocovariance lag window that is used to smooth a time series in the temporal dimension. The autocovariance lag window is determined by its weighting function and this weighting function depends on the number of Fourier frequencies, $k$, and the selected truncation point, $M$, such that $(2M + 1)$ is the width of the autocovariance lag window. Interested readers can see Koopmans (1995) and Wei (2006) for thoroughgoing discussions of autocovariance lag windows and their related spectral smoothing windows.

In the confidence interval of Equation (1.1.4), if $\omega$ is not at a Fourier frequency or if a smoothing window does not utilize uniform weights, as is the case with static smoothing windows, then degrees of freedom, $\nu$, can only be estimated and the equivalent degrees of freedom for such windows is:

$$\nu = \frac{2n}{M\int_{-1}^{1} W^2(x)dx} \quad (1.1.9)$$

where:

$n$ = number of observations;
$M$ = truncation point of the respective autocorrelation lag window in the temporal dimension; and
$W(x)$ = continuous weighting function for the window.

Typical static smoothing windows used in spectral analysis are the rectangular window (Wei, 2006), Tukey-Hamming and Tukey-Hanning windows (Blackman & Tukey, 1958a, 1958b), Bartlett's window (Bartlett, 1950), and Parzen's window (Parzen, 1961, 1963); the respective degrees of freedom for each of these is: $n/M$, $2.5n/M$, $2.67n/M$, $3n/M$, and $3.7n/M$ (Koopmans, 1995; Priestley, 1981; Wei, 2006). Thus, for a given static smoothing window, a given number of observations, $n$, and the selected truncation point, $M$, the width of its confidence interval is constant across all frequencies in the spectrum.

However, when the smoothing window utilizes uniform weights, as is the case of the simple moving average used in dynamic smoothing (i.e., DiRienzo-Zurbenko algorithm, Neagu-Zurbenko algorithm), its degrees of freedom is simply

$$\nu = (2m + 1) * 2 = 4m + 2 \quad (1.1.10)$$

where $m$ is the truncation point of the spectral smoothing window, $2m + 1$ is the spectral smoothing window width, and the factor 2 is the degrees of freedom for each Fourier frequency utilized within the smoothing window of width $2m + 1$.

The width of this window depends on the amount of variance (for the DiRienzo-Zurbenko algorithm) or the amount of departure from linearity (for the Neagu-Zurbenko algorithm) in the neighborhood of the given frequency in addition to the analyst's selected *proportion of smoothness*. When a signal is present, the amount of variance or the departure from linearity increases, the width of the smoothing window (i.e., $2m + 1$) decreases, the degrees of freedom decreases, and the width of the confidence interval is greatest. Thus, when a signal is present, the Kolmogorov-Zurbenko periodogram is least precise in its estimate of signal strength. As a result, it is important to compare the performance of the Kolmogorov-Zurbenko periodogram with dynamic smoothing windows to log-periodograms with static smoothing windows in the presence of a signal.

## 1.2 Candidate Functions to Compute and Plot Confidence Intervals for Dynamic Smoothing

Instructions for accessing and utilizing the *kza* package on the *R* platform to compute and plot standard Kolmogorov-Zurbenko periodograms with dynamic smoothing can be found in a paper by Loneck and colleagues (2024). However, computation of confidence intervals and their inclusion on plots required modification of two of its functions. Consequently, two candidate functions, *smoothwCIs.kzp( )* and *plotwCIs.kzp( )* were developed by augmenting its original *smooth.kzp( )* and *plot.kzp( )* functions. The *R* code for *smoothwCIs.kzp( )* and *plotwCIs.kzp( )*, including Notes regarding changes to the original functions, can be found in Appendices 1 and 2, respectively.

Generating Kolmogorov-Zurbenko periodograms with either DiRienzo-Zurbenko algorithm or Neagu-Zurbenko algorithm smoothing that includes $(1-\alpha)*100\%$ confidence intervals utilizes four functions:

**totalSignal<-kzp(y, m, k)**
**kzPeriodogram<-smoothWithCIs.kzp(totalSignal, log=T, smooth_level, method, alpha)**
**kzpPlot <- plotWithCIs.kzp(kzPeriodogram,type = "1")**
**kzpSummary <- summary.kzp(kzPeriodogram,digits,top)**

with respective arguments:

| | |
|---|---|
| y | raw time series dataset |
| m | initial window width of algorithm, $m_{Initial}$ |
| k | number of iterations of KZFT |
| smooth_level | proportion of total variance (for method = "DZ") *or* proportion of total departure from linearity (for method = "NZ") (a.k.a. proportion of smoothness) |
| method | method for smoothing: "DZ" for DiRienzo-Zurbenko, "NZ" for Neagu-Zurbenko |
| alpha | $\alpha$ used to determine $(1-\propto)*100\%$ confidence interval |
| digits | number of significant digits |
| top | number of top frequencies and periods returned |

To summarize, *kzp( )* generates raw periodogram ordinates; *smoothWithCIs.kzp( )* computes smoothed periodogram ordinates, confidence interval upper limits, and confidence interval lower limits; *plotWithCIs.kzp( )* generates a plot of the Kolmogorov-Zurbenko periodogram that is *mean-centered* and includes lower and upper confidence intervals; and *summary.kzp( )* provides the *top* frequencies and periods of interest.

Because the basic *plot.kzp( )* function uses *mean-centered* ordinates, the ordinates of the confidence intervals lower limit values and the ordinates of the confidence interval upper limit values are, likewise, adjusted by the same amount. More specifically, the ordinates of the confidence interval lower limit values and the ordinates of the confidence interval upper limit values are adjusted by subtracting the mean of the Kolmogorov-Zurbenko periodogram ordinates.

## 1.3 Demonstration of Confidence Intervals for Dynamic Smoothing

To illustrates the use of confidence intervals with Kolmogorov-Zurbenko periodograms with dynamic smoothing, four scenarios were considered. The first two are of a single signal embedded in a high level of noise, where one scenario uses DiRienzo-Zurbenko algorithm smoothing and the other uses Neagu-Zurbenko algorithm smoothing. The second two scenarios are with pairs of signals, close in frequency – one relatively strong and the other relatively weak – also embedded in a high level of noise, again where one scenario uses the DiRienzo-Zurbenko algorithm and the other uses the Neagu-Zurbenko algorithm.

**1.3.1 Illustrations: One Signal**

A simulated dataset was generated in which a single signal was embedded in a high level of noise. More specifically, a time series dataset of $n = 5000$ observations was created such that a signal frequency of $f = 0.444$ with signal amplitude $a_S = 3.58$ was embedded in random noise with amplitude $a_N = 16$, resulting in a signal-to-noise ratio of $0.050$. Next, a Kolmogorov-Zurbenko periodogram with DiRienzo-Zurbenko algorithm smoothing was constructed with proportion of smoothness $DZ = 0.05$, an initial window width of $m_{Initial} = 500$, and a 95% confidence interval for signal strength. Thus, this Kolmogorov-Zurbenko periodogram with DiRienzo-Zurbenko algorithm smoothing had sufficient sensitivity to detect a signal with a signal-to-noise ratio greater than or equal to 0.05 and could estimate a frequency with an accuracy of 0.002.

The resulting Kolmogorov-Zurbenko periodogram with DiRienzo-Zurbenko algorithm smoothing is shown in Figure 1.3.1.1. The frequency of the signal is clearly identified at exactly $\hat{f} = 0.444$ and the point estimate for the amplitude is $\widehat{a_S} = 3.425$ with a 95% confidence interval of $[1.783, 21.527]$. In addition, one can see that the width of the confidence interval is relatively narrow when no signal is present, fans out wider in approaching the known frequency ($f = 0.444$), achieves maximum width at that known frequency, and then contracts when moving away from the known frequency. This is so because confidence interval width is inversely proportional to degrees of freedom which are, in turn, proportional to the smoothing window width, and smoothing window width is smallest when a signal is present.

The simulation was repeated but with the Kolmogorov-Zurbenko periodogram constructed using Neagu-Zurbenko algorithm smoothing, with proportion of smoothness set at $NZ = 0.05$, an initial window width of $m_{Initial} = 500$, and a 95% confidence interval for signal strength. Thus, this Kolmogorov-Zurbenko periodogram with Neagu-Zurbenko algorithm smoothing had sufficient sensitivity to detect a signal with a signal-to-noise ratio greater than or equal to 0.05 and could estimate a frequency with an accuracy of 0.002.

The resulting periodogram is shown in Figure 1.3.1.2. Again, the frequency of the signal is clearly identified at $\hat{f} = 0.444$ and the point estimate for the amplitude is $\widehat{a_S} = 1.740$, with a 95% confidence interval of $[1.121, 3.832]$. As with DiRienzo-Zurbenko algorithm smoothing, the confidence interval is wide when a signal is present and narrows when no signal is present. Nevertheless, the width of the confidence interval when using the Neagu-Zurbenko algorithm was substantially narrower than the width of the confidence interval when using the DiRienzo-Zurbenko algorithm, indicating greater precision when only one signal is present.

One will note that the Kolmogorov-Zurbenko periodogram with DiRienzo-Zurbenko algorithm smoothing is relatively smooth, while the Kolmogorov-Zurbenko periodogram with Neagu-Zurbenko algorithm smoothing is jagged. This owes to the fact that DiRienzo-Zurbenko smoothing is based on proportion of total variance, while Neagu-Zurbenko smoothing is based on proportion of total departure from linearity. It has been noted that Neagu-Zurbenko algorithm smoothing can be more sensitive to signals than DiRienzo-Zurbenko algorithm smoothing. Consequently, the Kolmogorov-Zurbenko periodogram with Neagu-Zurbenko algorithm smoothing "expends" some of this sensitivity on random noise and, as a result, its ordinate in the presence of a signal frequency is lower than that of the Kolmogorov-Zurbenko periodogram with DiRienzo-Zurbenko algorithm smoothing.

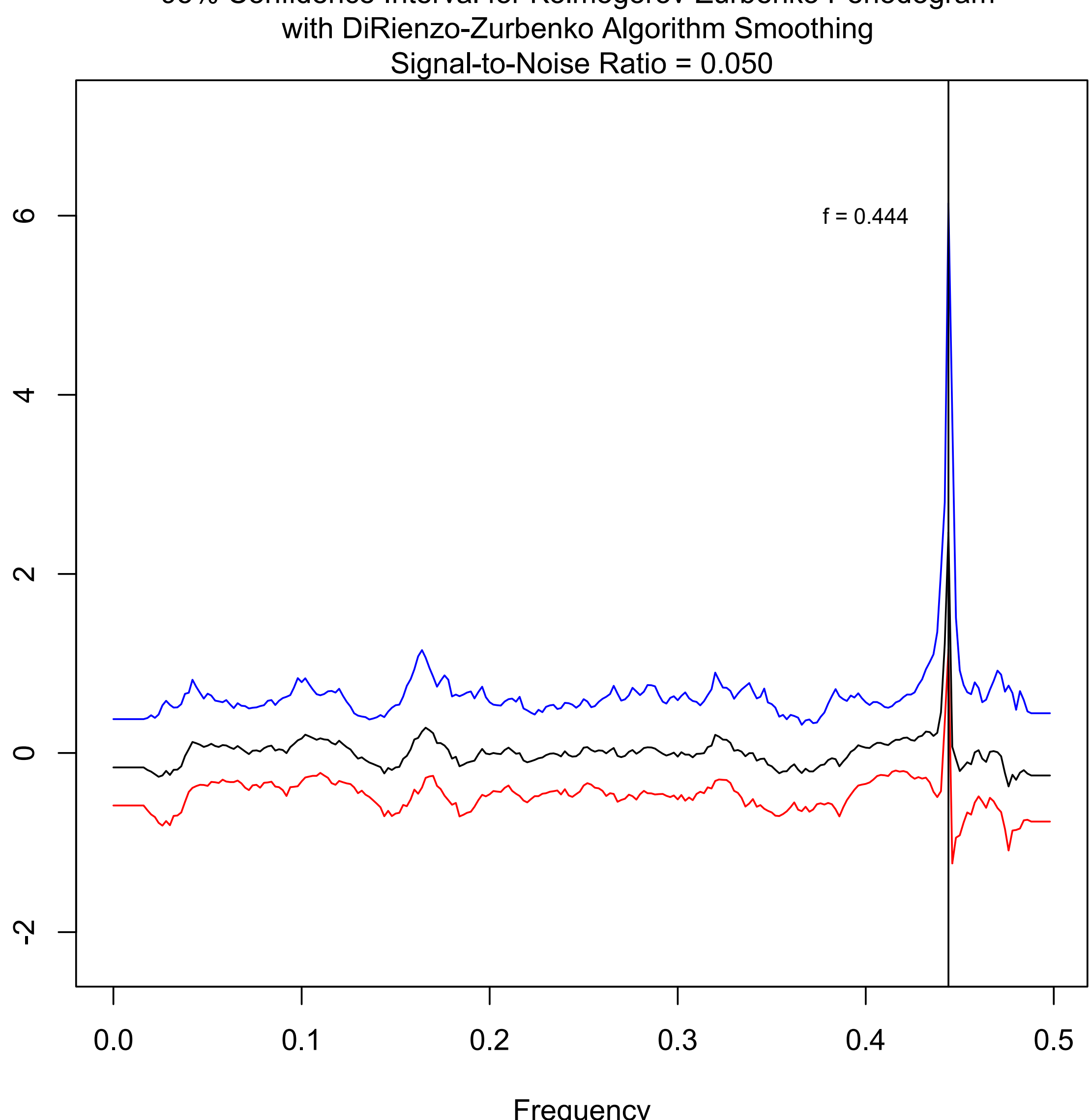

**FIGURE 1.3.1.1. 95% Confidence Interval for Kolmogorov-Zurbenko Periodogram Using DiRienzo-Zurbenko Algorithm Smoothing to Detect a Signal at Frequency $f = 0.444$ and Signal-to-Noise Ratio $= 0.05$.**

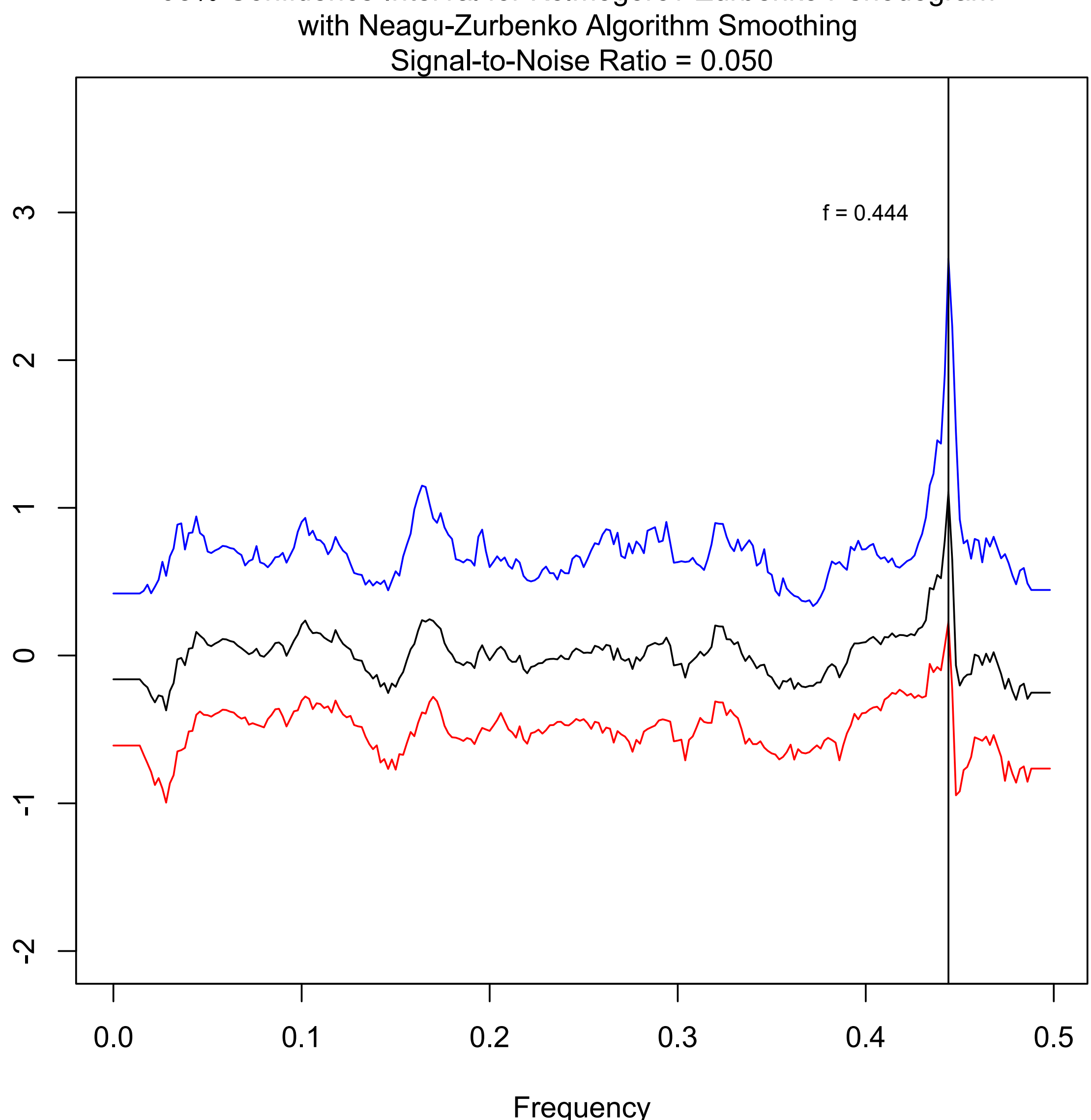

**FIGURE 1.3.1.2. 95% Confidence Interval for Kolmogorov-Zurbenko Periodogram Using Neagu-Zurbenko Algorithm Smoothing to Detect a Signal at Frequency $f = 0.444$ and Signal-to-Noise Ratio $= 0.05$.**

In sum, the Kolmogorov-Zurbenko periodogram with DiRienzo-Zurbenko algorithm smoothing had a point estimate of amplitude that was quite close to the actual value (i.e., 3.425 versus 3.59) and its 95% confidence interval (i.e., of $[1.783, 21.527]$) included that actual value. However, when the Neagu-Zurbenko algorithm was used, the point estimate for the amplitude was at a greater distance from the actual value (1.740 versus 3.59); nevertheless, it's 95% confidence interval (i.e., $[1.121, 3.832]$) was narrower, yet still included the actual value.

**1.3.2 Illustration: Two Signals**

A second simulated dataset was generated in which two signals, close in frequency, one relatively strong and one relatively weak, were embedded in a high level of noise. More specifically, a time series dataset of $n = 5000$ observations was created such that one signal had a frequency of $f_1 = 0.400$ with an amplitude of $a_{S_1} = 8$ and a second signal had a frequency of $f_2 = 0.380$ with an amplitude of $a_{S_2} = 4$, both of which were embedded in random noise with an amplitude of $a_N = 16$, resulting in a signal-to-noise ratio of $0.3125$. A Kolmogorov-Zurbenko periodogram with DiRienzo-Zurbenko algorithm smoothing was constructed with proportion of smoothness $DZ = 0.05$, an initial window width of $m_{Initial} = 500$, and a 95% confidence interval. Thus, the Kolmogorov-Zurbenko periodogram with DiRienzo-Zurbenko algorithm smoothing had sufficient sensitivity to detect signals with total signal-to-noise ratios greater than or equal to 0.05, could estimate frequencies with an accuracy within 0.002, and could resolve frequency pairs as close as 0.004.

The resulting Kolmogorov-Zurbenko periodogram with DiRienzo-Zurbenko algorithm smoothing is shown in Figure 1.3.2.1. Two signals were clearly identified. The frequency of the stronger signal was exactly $\widehat{f_1} = 0.400$ and the point estimate of its amplitude was $\widehat{a_{S_1}} = 7.290$, with a 95% confidence interval of $[3.796, 45.816]$; likewise, the frequency of the weaker signal was exactly $\widehat{f_2} = 0.380$ and a point estimate for its amplitude was $\widehat{a_{S_2}} = 3.647$, with a 95% confidence interval of $[1.899, 22.923]$.

As with the Kolmogorov-Zurbenko periodogram of a single signal frequency, the confidence interval width is narrow when no signal is present, fans out in approaching both signal frequencies, is at a local maxim width at the known signal frequencies, then contracts when moving away from the known signal frequencies. Again, this is so because confidence interval width is inversely proportional to degrees of freedom which are, in turn, proportional to the smoothing window width, and smoothing window width has a local minimum when a signal is present.

The simulation was repeated, but with the Kolmogorov-Zurbenko periodogram constructed using Neagu-Zurbenko algorithm smoothing, with proportion of smoothness set at $NZ = 0.05$, an initial window width of $m_{Initial} = 500$, and a 95% confidence interval. Thus, the Kolmogorov-Zurbenko periodogram with Neagu-Zurbenko algorithm smoothing had sufficient sensitivity to detect signals with total signal-to-noise ratios greater than or equal to 0.05, could estimate frequencies with an accuracy within 0.002, and could resolve frequency pairs as close as 0.004.

The resulting periodogram is shown in Figure 1.3.2.2. Again, two signals are clearly identified. The frequency of the stronger signal was exactly $\widehat{f_1} = 0.400$ and the point estimate for its amplitude was $\widehat{a_{S_1}} = 7.297$, with a 95% confidence interval of $[3.799, 45.863]$; likewise, the frequency of the weaker signal was $\widehat{f_2} = 0.380$ and the point estimate for its amplitude was $\widehat{a_{S_2}} = 3.651$, with a 95% confidence interval of $[1.901, 22.947]$. As when the DiRienzo-Zurbenko algorithm is used, the confidence interval is wide when a signal is present and narrow when no signal is present. The width of the Neagu-Zurbenko algorithm confidence intervals was commensurate with the width of the DiRienzo-Zurbenko algorithm confidence intervals, indicating similar levels of precision when two signals are present.

Although the Kolmogorov-Zurbenko periodogram with Neagu-Zurbenko algorithm smoothing is more jagged than when the DiRienzo-Zurbenko algorithm is used, the extent of this jaggedness is less pronounced than in the scenario with a single signal. This owes to the fact that the Neagu-Zurbenko algorithm expends more of its sensitivity on the second weaker signal, with less sensitivity left for the random noise. Consequently, when two signals are present, the jaggedness of Kolmogorov-Zurbenko periodogram with Neagu-Zurbenko algorithm smoothing is on par with that of the Kolmogorov-Zurbenko periodogram with DiRienzo-Zurbenko algorithm.

In sum, the Kolmogorov-Zurbenko periodogram with DiRienzo-Zurbenko algorithm smoothing had point estimates for the amplitude that were quite close to the actual values (7.290 versus 8.000 and 3.647 versus 4.000), with the respective 95% confidence intervals ([3.796, 45.816] and [1.899, 22.923]) including the actual values (8.00 and 4.00). Similarly, when the Neagu-Zurbenko algorithm was used, its point estimates for the amplitudes were also quite close to the actual values (7.297 versus 8.000 and 3.651 versus 4.00), with the respective 95% confidence intervals ([3.799, 45.863] and [1.901, 22.947]) including the respective actual frequency values (8.000 and 4.000).

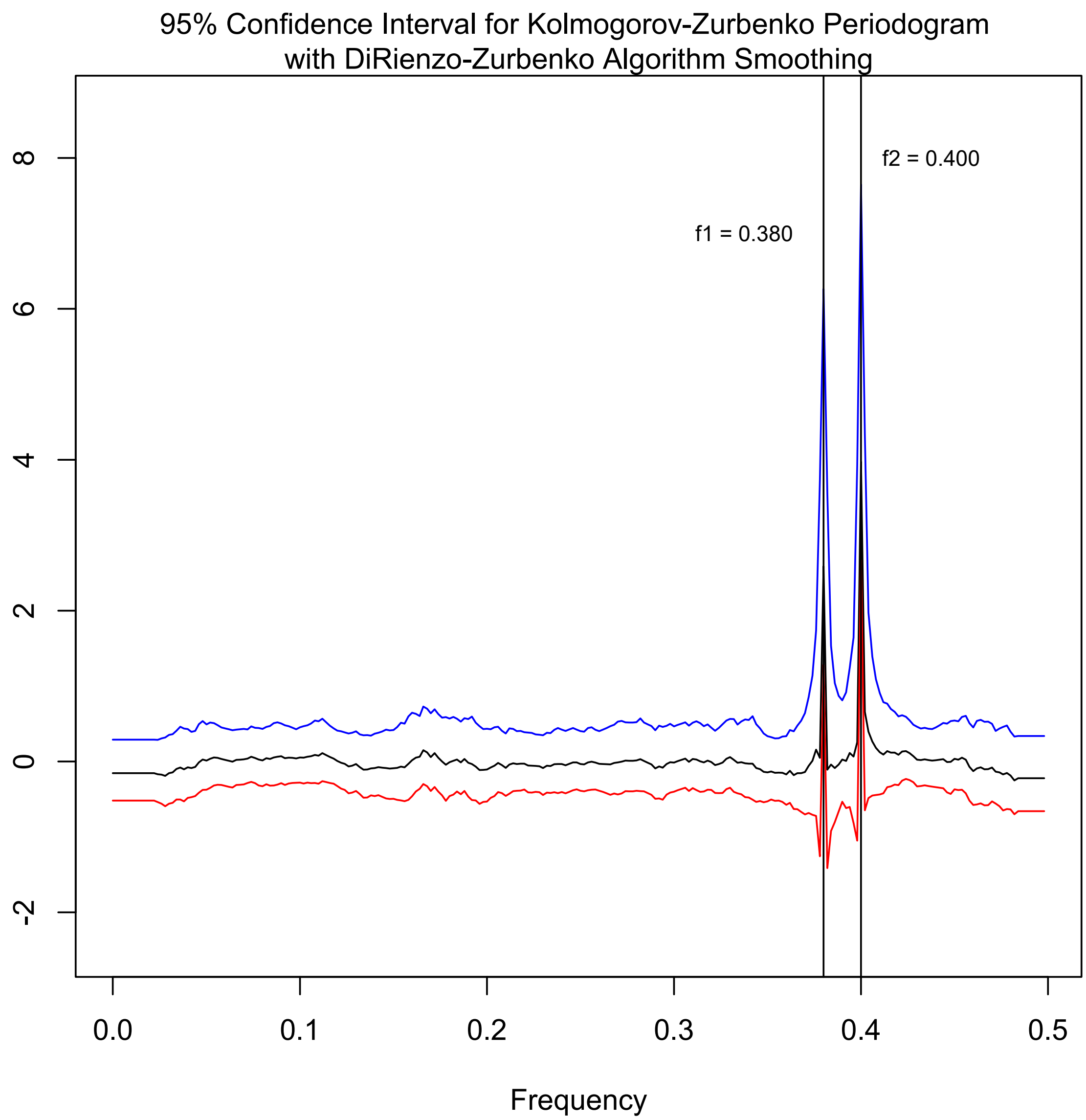

**FIGURE 1.3.2.1. 95% Confidence Interval for Kolmogorov-Zurbenko Periodogram using DiRienzo-Zurbenko Algorithm Smoothing to Resolve Frequencies at $f_1 = 0.380$ and $f_2 = 0.400$ with Total Signal-to-Noise Ratio $= 0.3125$.**

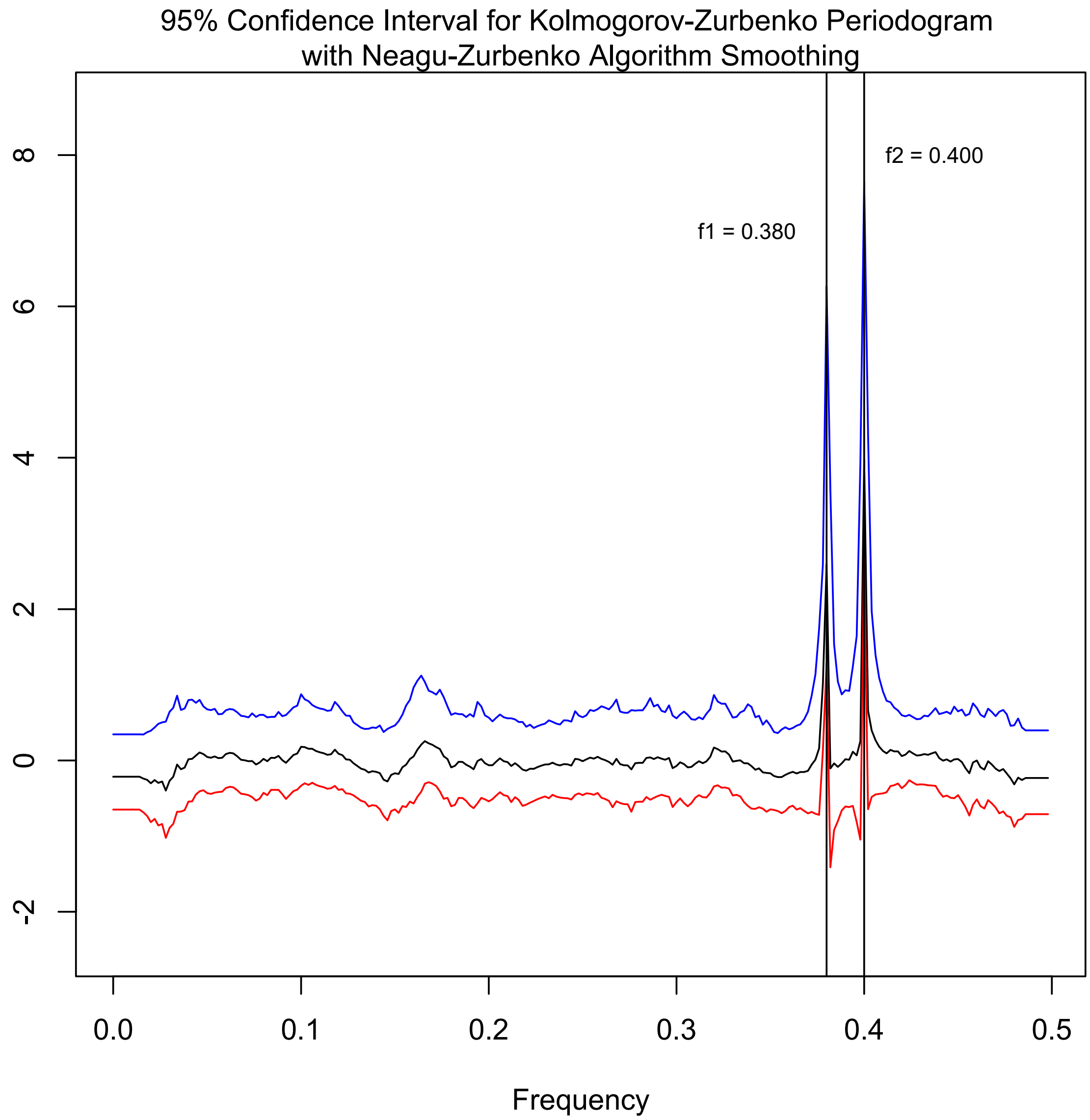

**FIGURE 1.3.2.2. 95% Confidence Interval for Kolmogorov-Zurbenko Periodogram using Neagu-Zurbenko Algorithm Smoothing to Resolve Frequencies at $f_1 = 0.380$ and $f_2 = 0.400$ with Total Signal-to-Noise Ratio $= 0.3125$.**

**2. Precision of Signal Strength Estimates**

To establish and compare the theoretical and practical limits of signal strength estimate precision for Kolmogorov-Zurbenko periodograms with dynamic smoothing to standard log-periodograms with static smoothing, this investigation contrasted their respective confidence interval widths. This section provides: a brief definition of signal strength estimate precision, its importance, and related considerations; theoretical bases for comparing the confidence intervals of Kolmogorov-Zurbenko periodograms with dynamic smoothing to log-periodograms with static smoothing, along with a comparison of their theoretical and practical limits in the presence of a signal; and illustrations contrasting the respective confidence interval widths.

**2.1 Definition, Importance, and Considerations**

The precision of signal strength estimates is the ability to accurately estimate the amplitude of a signal at a given frequency. Previous research assessed Kolmogorov-Zurbenko periodograms with dynamic smoothing with respect to signal frequency estimates using four criteria: sensitivity, accuracy, resolution, and robustness with missing data (Loneck et al., 2024; Zurbenko et al., 2020). However, Kolmogorov-Zurbenko periodograms not only estimate signal frequencies, but also estimate signal strength. While periodogram sensitivity, accuracy, and resolution are necessary conditions for detecting, identifying, and separating signal frequencies, they are not sufficient for framing a model to predict future values of a variable. In addition, a periodogram must be able to estimate signal strength (i.e., amplitude) with adequate precision. This issue is important because both signal frequencies and their respective signal amplitudes are needed to provide a comprehensive estimate for future values of a periodic time series. Consequently, signal strength estimate precision is a fifth criterion for evaluating the Kolmogorov-Zurbenko periodogram.

Because random noise is always present, precision of a periodogram's estimated signal strength at a given frequency is provided by the confidence interval for the signal strength at that frequency – the narrower the confidence interval, the greater the precision, and the wider the confidence interval, the less the precision. Thus, confidence intervals for Kolmogorov-Zurbenko periodograms with dynamic smoothing can be used to establish the precision with which it estimates signal strength by comparing the theoretical and respective practical limits of their confidence interval widths to the theoretical and respective practical limits of confidence interval width for static smoothing, in the presence of a signal.

With dynamic smoothing windows, Section 1.1 pointed out that its confidence interval width varies and this variation depends on the smoothing window width at a given frequency; the window width, in turn, depends on the presence or absence of a signal at that frequency. When a signal is present, the dynamic smoothing window shrinks and the confidence interval is at maximum width; when a signal is not present, the smoothing window expands and the confidence interval width is at a minimum. It must be remembered that $m$ is the truncation point of the dynamic smoothing window on either side of the given frequency and $window\ width = 2m + 1$, with $m \geq 1$. Consequently, when a signal is present, its window width can be as small as 3, its degrees of freedom, $4m + 2$, are at a minimum, and the confidence interval is at maximum width. Conversely, when no signal is present, window width is at a maximum and, theoretically, can be as wide as $PoS * n$, with $PoS =$ proportion of smoothness and $n =$ number of observations in the time series; it is here that its degrees of freedom are at a maximum and the confidence interval width is at a minimum.

With static smoothing windows, Section 1.1 also pointed out that their confidence intervals are determined by their respective degrees of freedom, $\nu$, which are, in turn determined by $M$, the truncation point of the respective autocorrelation lag window as selected by the analyst, and $n$, the number of observations in the time series. Because window width is constant across the spectrum of Fourier frequencies, its confidence interval is, likewise, constant across the spectrum.

## 2.2 Statistical Bases and Limits

Given an identified signal frequency, a static smoothing window and its smoothing window width can be selected such that its estimate of signal strength is more precise than that of a dynamic smoothing window. Based on Equation 1.1.4 and for a $(1-\alpha)$ percent level of confidence, the width of a confidence interval for a log-periodogram is:

$$CI\,Width = ln\left[\frac{\nu}{\chi^2_{1-\alpha/2}(\nu)}\right] - ln\left[\frac{\nu}{\chi^2_{\alpha/2}(\nu)}\right] \tag{2.2.1}$$

which can be simplified to:

$$CI\,Width = ln\left(\frac{\chi^2_{\alpha/2}(\nu)}{\chi^2_{1-\alpha/2}(\nu)}\right) \tag{2.2.2}$$

From Equation 2.2.2, it is clear that the larger the respective degrees of freedom, the smaller the confidence interval and, as stated earlier, the smaller the confidence interval, the more precise the estimate of signal strength.

To consider the theoretical limit of dynamic smoothing windows, the focus turns to what occurs when a signal is present because this is where the confidence interval width of a dynamic window is critical in determining the precision of its signal strength estimate. When a signal is present, the theoretical minimum window width for a dynamic smoothing window is 3, indicating a truncation point of $m = 1$ and, thus, $\nu = 4m + 2 = 6$ degrees of freedom. This results in a confidence interval width of

$$CI\,Width_{Dynamic} = ln\left(\frac{\chi^2_{\alpha/2}(\nu=6)}{\chi^2_{1-\alpha/2}(\nu=6)}\right) \tag{2.2.3}$$

As noted in Section 1.1, the specific static windows under consideration and their respective degrees of freedom are rectangular with $\nu = n/M$, Tukey-Hamming with $\nu = 2.5n/M$; Tukey-Hanning with $\nu = 2.67n/M$; Bartlett with $\nu = 3n/M$, and Parzen with $\nu = 3.7n/M$. Thus, for a given number of observations, $n$, analysts can ensure a more precise signal strength estimate for a given static smoothing window by selecting a respective static smoothing window width with truncation point $M$ such that its respective degrees of freedom are greater than the theoretical limit for dynamic smoothing, $\nu = 6$.

For each static smoothing window under consideration, this results in an upper limit for its confidence interval width with an implicit upper limit of its truncation point such that it produces a more precise estimate of signal strength. These are:

$$CI\ Width_{Rectangular} = ln\left(\frac{\chi^2_{\alpha/2}(v=\frac{n}{M}>6)}{\chi^2_{1-\alpha/2}(v=\frac{n}{M}>6)}\right) \Rightarrow M < \frac{1}{6}n \quad (2.2.4a)$$

$$CI\ Width_{Tukey-Hamming} = ln\left(\frac{\chi^2_{\alpha/2}(v=\frac{2.5n}{M}>6)}{\chi^2_{1-\alpha/2}(v=\frac{2.5n}{M}>6)}\right) \Rightarrow M < \frac{2.5}{6}n \quad (2.2.4b)$$

$$CI\ Width_{Tukey-Hanning} = ln\left(\frac{\chi^2_{\alpha/2}(v=\frac{2.67n}{M}>6)}{\chi^2_{1-\alpha/2}(v=\frac{2.67n}{M}>6)}\right) \Rightarrow M < \frac{2.67}{6}n \quad (2.2.4c)$$

$$CI\ Width_{Bartlett} = ln\left(\frac{\chi^2_{\alpha/2}(v=\frac{3n}{M}>6)}{\chi^2_{1-\alpha/2}(v=\frac{3n}{M}>6)}\right) \Rightarrow M < \frac{3}{6}n \quad (2.2.4d)$$

$$CI\ Width_{Parzen} = ln\left(\frac{\chi^2_{\alpha/2}(v=\frac{3.7n}{M}>6)}{\chi^2_{1-\alpha/2}(v=\frac{3.7n}{M}>6)}\right) \Rightarrow M < \frac{3.7}{6}n \quad (2.2.4e)$$

As a practical matter, an analyst must also ensure that the static smoothing window can resolve signals close in frequency at least as well as the dynamic smoothing window and this leads to identifying the lower bound for truncation point $M$. Wei (2006) posited an overarching principal that $M$ should be selected in such a way that the bandwidth of the respective static smoothing window "... should not exceed the minimum interval between adjacent peaks ..." on the periodogram (p. 313). The estimated bandwidths of the static smoothing windows under consideration for a given truncation point, $M$, are: rectangular with $\pi/M$; Tukey-Hamming with $2.5\pi/M$; Tukey-Hanning with $2.67\pi/M$; Bartlett with $3\pi/M$; and Parzen with $3.7\pi/M$ (Koopmans, 1995). Thus, for a given static smoothing window, analysts can select a truncation point such that its resolution is a good as or better than that of a dynamic smoothing window.

If the closest two frequencies identified with a Kolmogorov-Zurbenko periodogram with dynamic smoothing are $\lambda_i$ and $\lambda_{i+1}$, then the analyst must select $M$ such that:

$$static\ smoothing\ bandwith\ \leq\ |\lambda_{i+1} - \lambda_i| \quad (2.2.5)$$

Thus, the resulting inequalities and implicit lower bounds for the truncation points, $M$, of the static windows under consideration are:

Rectangular: $\frac{\pi}{M} \leq |\lambda_{i+1} - \lambda_i| \Rightarrow \frac{\pi}{|\lambda_{i+1}-\lambda_i|} \leq M$ (2.2.6a)

Tukey-Hamming: $\frac{2.5\pi}{M} \leq |\lambda_{i+1} - \lambda_i| \Rightarrow \frac{2.5\pi}{|\lambda_{i+1}-\lambda_i|} \leq M$ (2.2.6b)

Tukey-Hanning: $\frac{2.67\pi}{M} \leq |\lambda_{i+1} - \lambda_i| \Rightarrow \frac{2.67\pi}{|\lambda_{i+1}-\lambda_i|} \leq M$ (2.2.6c)

Bartlett: $\frac{3\pi}{M} \leq |\lambda_{i+1} - \lambda_i| \Rightarrow \frac{3\pi}{|\lambda_{i+1}-\lambda_i|} \leq M$ (2.2.6d)

Parzen: $\frac{3.7\pi}{M} \leq |\lambda_{i+1} - \lambda_i| \Rightarrow \frac{3.7\pi}{|\lambda_{i+1}-\lambda_i|} \leq M$ (2.2.6e)

The lower bound of M achieves a minimum when $|\lambda_{i+1} - \lambda_i|$ spans $\pi$ (i.e., when $\frac{\pi}{|\lambda_{i+1}-\lambda_i|} = 1$). Thus, the minimum lower bound for each static smoothing window is: 1 for rectangular, 2.5 for Tukey-Hamming, 2.67 for Tukey-Hanning, 3 for Bartlett, and 3.7 for Parzen. These values are needed when comparing the confidence interval widths of dynamic smoothing windows to those of static smoothing windows and will be used in the illustrations presented in the next subsection.

The lower bound and the upper bound of truncation point $M$ for each of the static smoothing windows is summarized in Table 2.2.1. Upon inspection, one can see that there is a common constraint on the difference in the two closest frequencies identified by a Kolmogorov-Zurbenko periodogram with dynamic smoothing if the static smoothing window is to resolve frequencies as well as a dynamic smoothing window. Clearly, the lower bound must be less than the upper bound and this results in an inequality for each static smoothing window that leads to an implicit constraint on $|\lambda_{i+1} - \lambda_i|$. These are:

Rectangular: $\frac{\pi}{|\lambda_{i+1}-\lambda_i|} < \frac{1}{6}n \Rightarrow \frac{6\pi}{n} < |\lambda_{i+1} - \lambda_i|$ (2.2.7a)

Tukey-Hamming: $\frac{2.5\pi}{|\lambda_{i+1}-\lambda_i|} < \frac{2.5}{6}n \Rightarrow \frac{6\pi}{n} < |\lambda_{i+1} - \lambda_i|$ (2.2.7b)

Tukey-Hanning: $\frac{2.67\pi}{|\lambda_{i+1}-\lambda_i|} < \frac{2.67}{6}n \Rightarrow \frac{6\pi}{n} < |\lambda_{i+1} - \lambda_i|$ (2.2.7c)

Bartlett: $\frac{3\pi}{|\lambda_{i+1}-\lambda_i|} < \frac{3}{6}n \Rightarrow \frac{6\pi}{n} < |\lambda_{i+1} - \lambda_i|$ (2.2.7d)

Parzen: $\frac{3.7\pi}{|\lambda_{i+1}-\lambda_i|} < \frac{3.7}{6}n \Rightarrow \frac{6\pi}{n} < |\lambda_{i+1} - \lambda_i|$ (2.2.7e)

The common implicit constraint on $|\lambda_{i+1} - \lambda_i|$ owes to the "static smoothing window constant" that serves as a factor in both the lower bound and upper bound of truncation point $M$.

In order for a static smoothing window's truncation point $M$ to lie between the lower bound as provided in Equations 2.2.6a through 2.2.6e and the upper bound as provided in Equations 2.2.4a through 2.2.4e, the difference between the two closest frequencies identified by a Kolmogorov-Zurbenko periodogram with dynamic smoothing must be greater than $\frac{6\pi}{n}$ so that precision exceeds that of the Kolmogorov-Zurbenko periodogram with dynamic smoothing, while maintaining adequate resolution to separate $\lambda_{i+1}$ and $\lambda_i$. However, if $|\lambda_{i+1} - \lambda_i|$ is less than or equal to $\frac{6\pi}{n}$, a log-periodogram with static smoothing will not be able to resolve the two frequencies and the analyst will have to rely on the Kolmogorov-Zurbenko periodogram with dynamic smoothing for the estimates of their respective signal strengths.

Based on the upper bounds and lower bounds of the static windows' truncation points, the next section compares the resulting confidence interval widths of static smoothing windows to those of dynamic smoothing windows across four scenarios. In each scenario, one sees that static smoothing windows can be used to obtain greater precision, along with comparable resolution, in estimating signal strength in contrast to the precision of dynamic smoothing windows, in the presence of a signal.

| STATIC SMOOTHING WINDOW | TRUNCATION POINT LOWER BOUND AND UPPER BOUND |
|---|---|
| **Rectangular** | $\frac{\pi}{\lvert\lambda_{i+1}-\lambda_i\rvert} \leq M < \frac{1}{6}n$ |
| **Tukey-Hamming** | $\frac{2.5\pi}{\lvert\lambda_{i+1}-\lambda_i\rvert} \leq M < \frac{2.5}{6}n$ |
| **Tukey-Hanning** | $\frac{2.67\pi}{\lvert\lambda_{i+1}-\lambda_i\rvert} \leq M < \frac{2.67}{6}n$ |
| **Bartlett** | $\frac{3\pi}{\lvert\lambda_{i+1}-\lambda_i\rvert} \leq M < \frac{3}{6}n$ |
| **Parzen** | $\frac{3.7\pi}{\lvert\lambda_{i+1}-\lambda_i\rvert} \leq M < \frac{3.7}{6}n$ |

**Table 2.2.1. Summary of Truncation Point Bounds for Signal Strength Precision and Signal Frequency Resolution Across Static Smoothing Windows.**

**2.3 Illustrations**
To illustrate the precision of the Kolmogorov-Zurbenko periodograms with dynamic smoothing with respect to signal strength, confidence interval widths for Kolmogorov-Zurbenko periodograms with dynamic smoothing were compared to confidence interval widths for log-periodograms with static smoothing across four scenarios. The first two assumed a time series with $n = 5000$ observations - one in which the dynamic smoothing window had a proportion of smoothness $PoS = 0.05$ and the window width ranged from 3 to 250 (i.e., $PoS * n = 0.05 * 5000$) and one in which the dynamic smoothing window had a proportion of smoothness $PoS = 0.01$ and the window width ranged from 3 to 50 (i.e., $PoS * n = 0.01 * 5000$). Similarly, the second two scenarios assumed a time series with $n = 1000$ – one in which the dynamic smoothing window had a proportion of smoothness $PoS = 0.05$ and the window width ranged from 3 to 50 (i.e., $PoS * n = 0.05 * 1000$) and one in which the dynamic smoothing window had a proportion of smoothness $PoS = 0.01$ and its window width ranged from 3 to 10 (i.e., $PoS * n = 0.01 * 1000$).

Within each scenario, the confidence interval widths of Kolmogorov-Zurbenko periodograms with dynamic smoothing were contrasted with confidence interval widths of log-periodograms with static smoothing, according to the conditional truncation point intervals established in Section 2.2. Thus, static smoothing confidence intervals were established for each static smoothing window under consideration at three separate values: largest static smoothing confidence interval (i.e., maximum truncation point), median static smoothing confidence interval (i.e., median truncation point), and smallest static smoothing confidence interval (i.e., minimum truncation point). Dynamic smoothing algorithms included the DiRienzo-Zurbenko algorithm and the Neagu-Zurbenko algorithm, which are equivalent with respect to confidence interval widths, while static smoothing algorithms included rectangular, Tukey-Hamming, Tukey-Hanning, Bartlett, and Parzen.

As noted earlier, precision, and thus confidence interval widths, are most critical when a signal is present and such is the case when the dynamic smoothing window width is at a minimum (i.e., dynamic smoothing window width is 3). Across all four scenarios, one sees that when a signal is present the confidence interval widths of static smoothing windows at their maximum truncation point are slightly smaller than confidence interval widths of dynamic smoothing windows; further, the confidence interval width of static smoothing windows grows narrower as it moves to the median and the minimum truncation point. See Figures 2.3.1. through Figure 2.3.4.

This indicates the truncation point of a given static smoothing window can be selected such that its confidence interval width is narrower than that of a dynamic smoothing window in the presence of a signal and, thus, can be made to be more precise in its estimate of signal strength. Further, if the difference in the closest two frequencies identified by a Kolmogorov-Zurbenko periodogram with dynamic smoothing is greater than $\frac{6\pi}{n}$, then resolution is as good as or better than that of a dynamic window.

Across all four scenarios, four additional trends were also evident, two with respect to dynamic smoothing and two with respect to static smoothing. With respect to dynamic smoothing, confidence interval widths are largest when the smoothing window is narrow (indicative of presence of a signal) and are smallest when the smoothing window is wide (indicative of absence of a signal). Thus, as noted earlier, precision decreases in the presence of a signal. In addition, the plot of confidence interval width over dynamic smoothing window width is the same across all four scenarios, with the only difference being the upper bound of the dynamic smoothing window width.

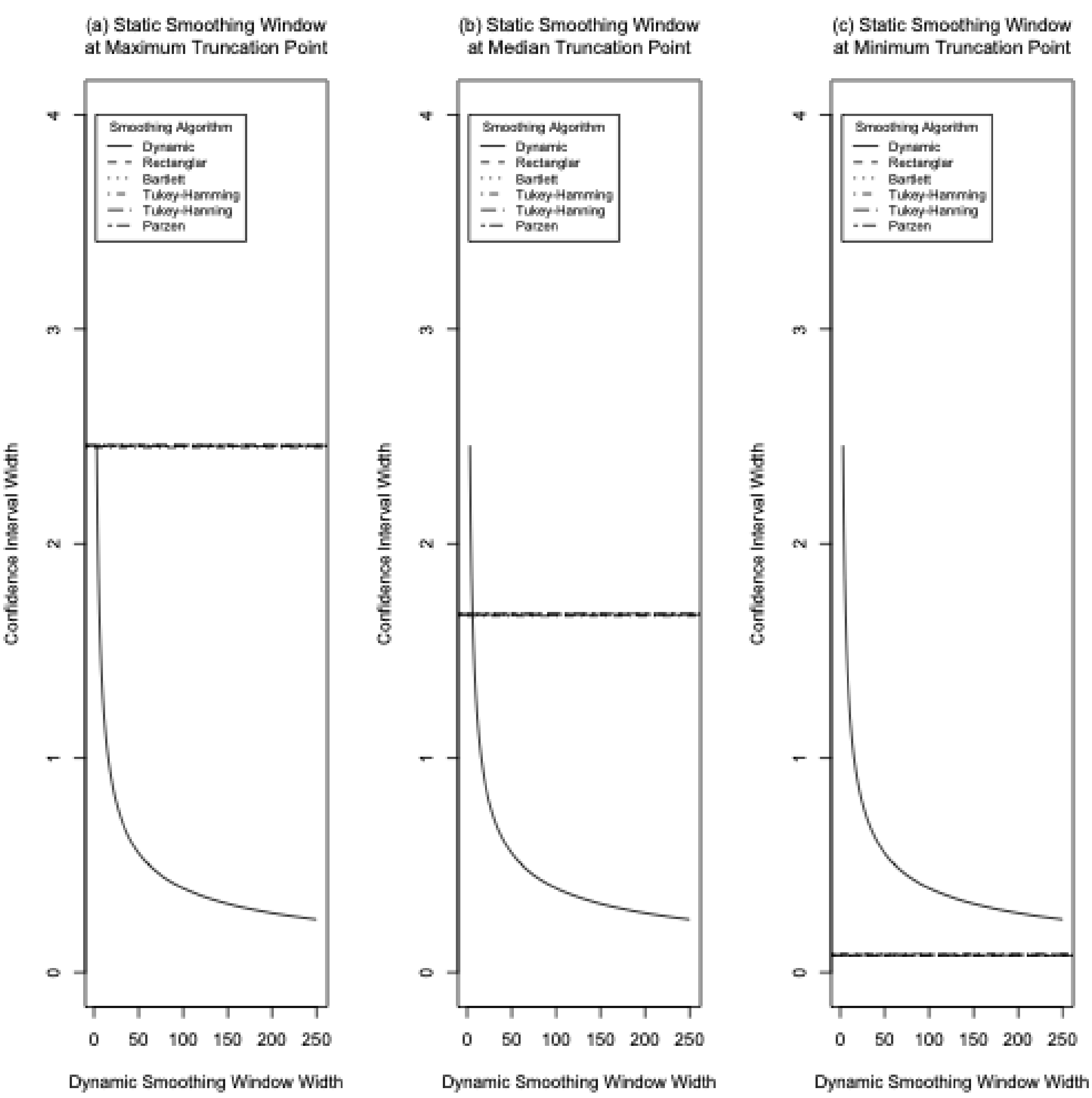

**FIGURE 2.3.1. A comparison of CI widths across possible dynamic smoothing window widths to CI widths for static smoothing window widths when $n = 5000$ and dynamic smoothing Proportion of Smoothing $= 0.05$.**

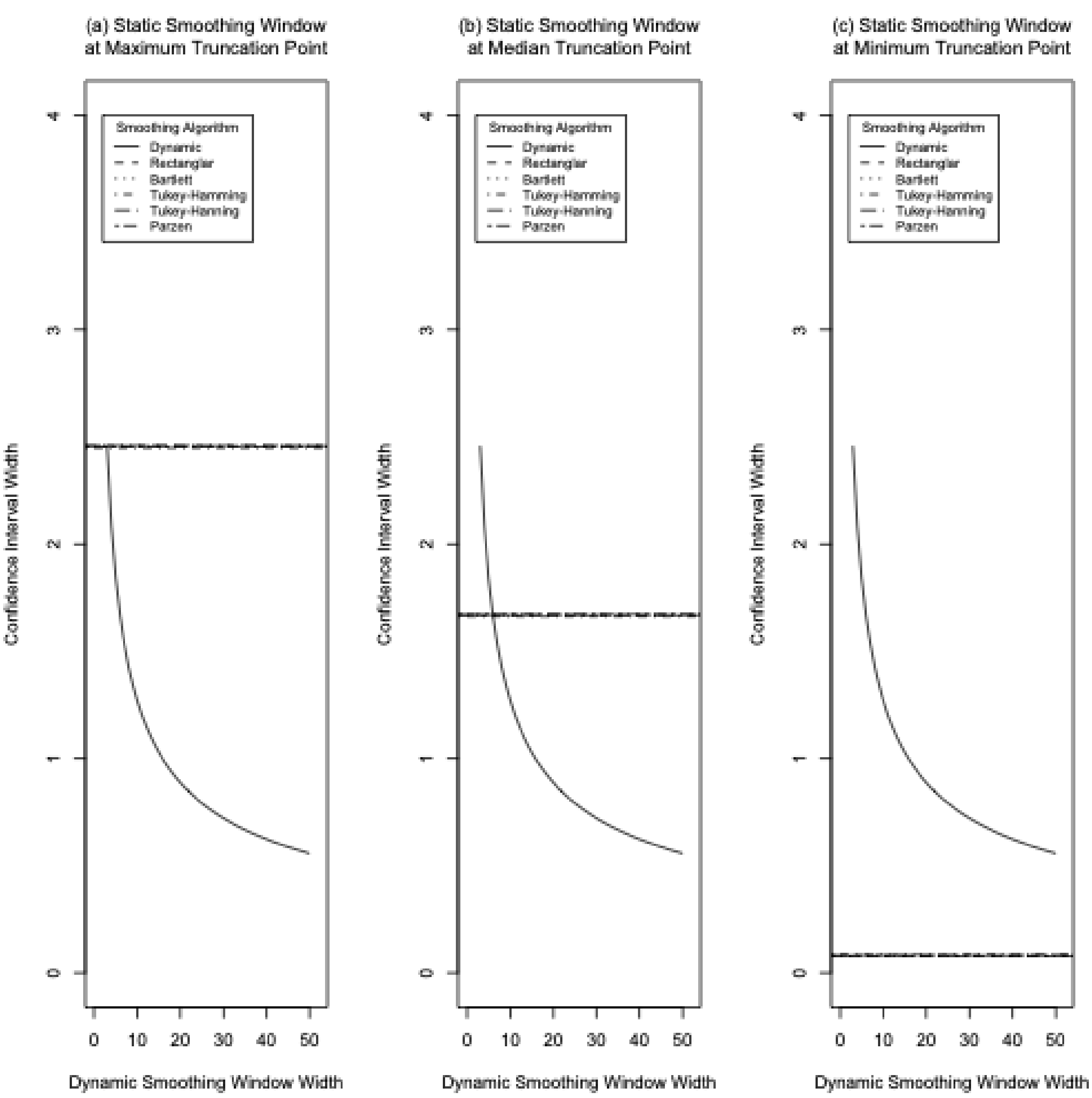

**FIGURE 2.3.2. A comparison of CI widths across possible dynamic smoothing window widths to CI widths for static smoothing window widths when $n = 5000$ and dynamic smoothing Proportion of Smoothing $= 0.01$.**

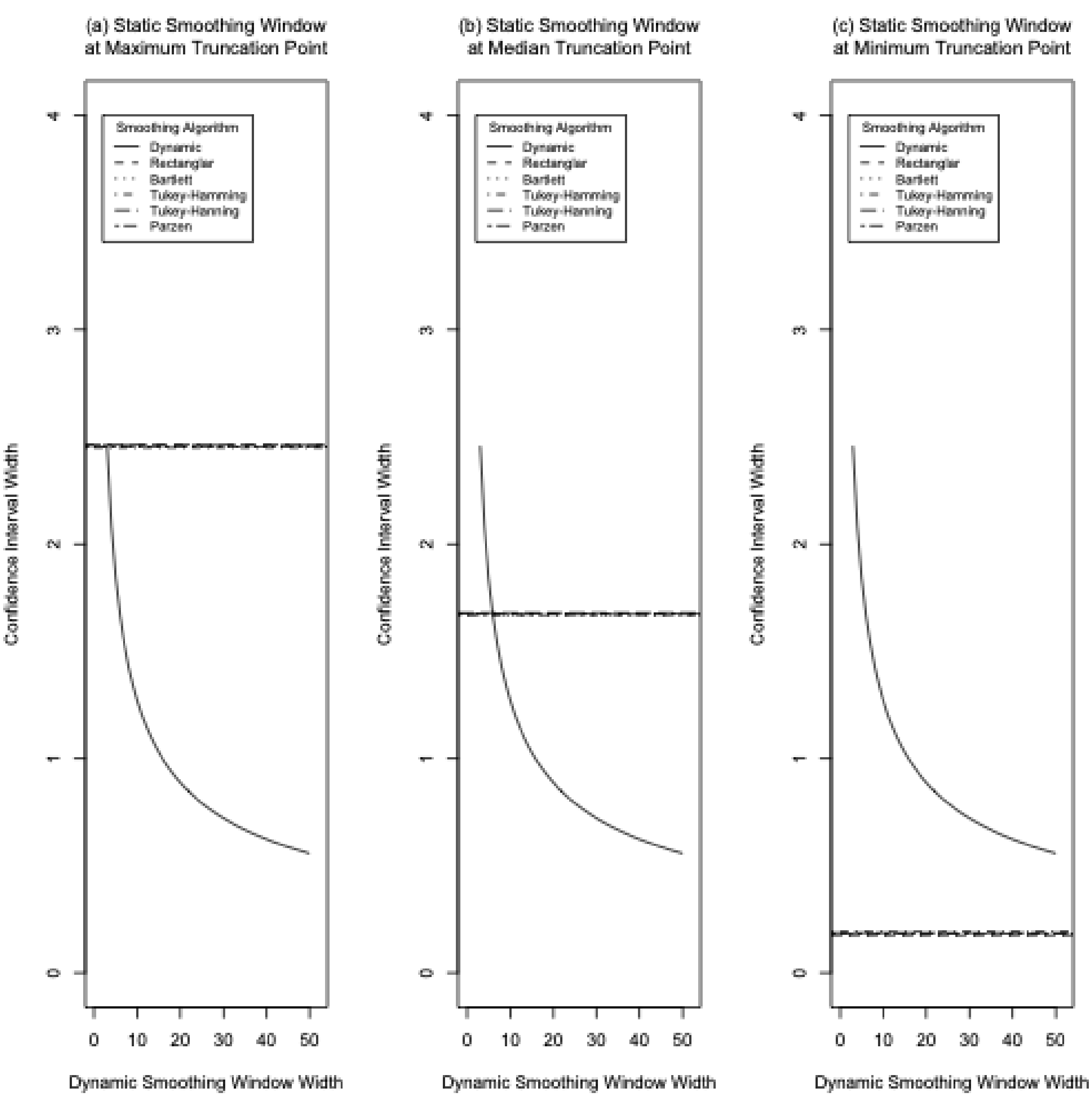

**FIGURE 2.3.3. A comparison of CI widths across possible dynamic smoothing window widths to CI widths for static smoothing window widths when $n = 1000$ and dynamic smoothing Proportion of Smoothing $= 0.05$.**

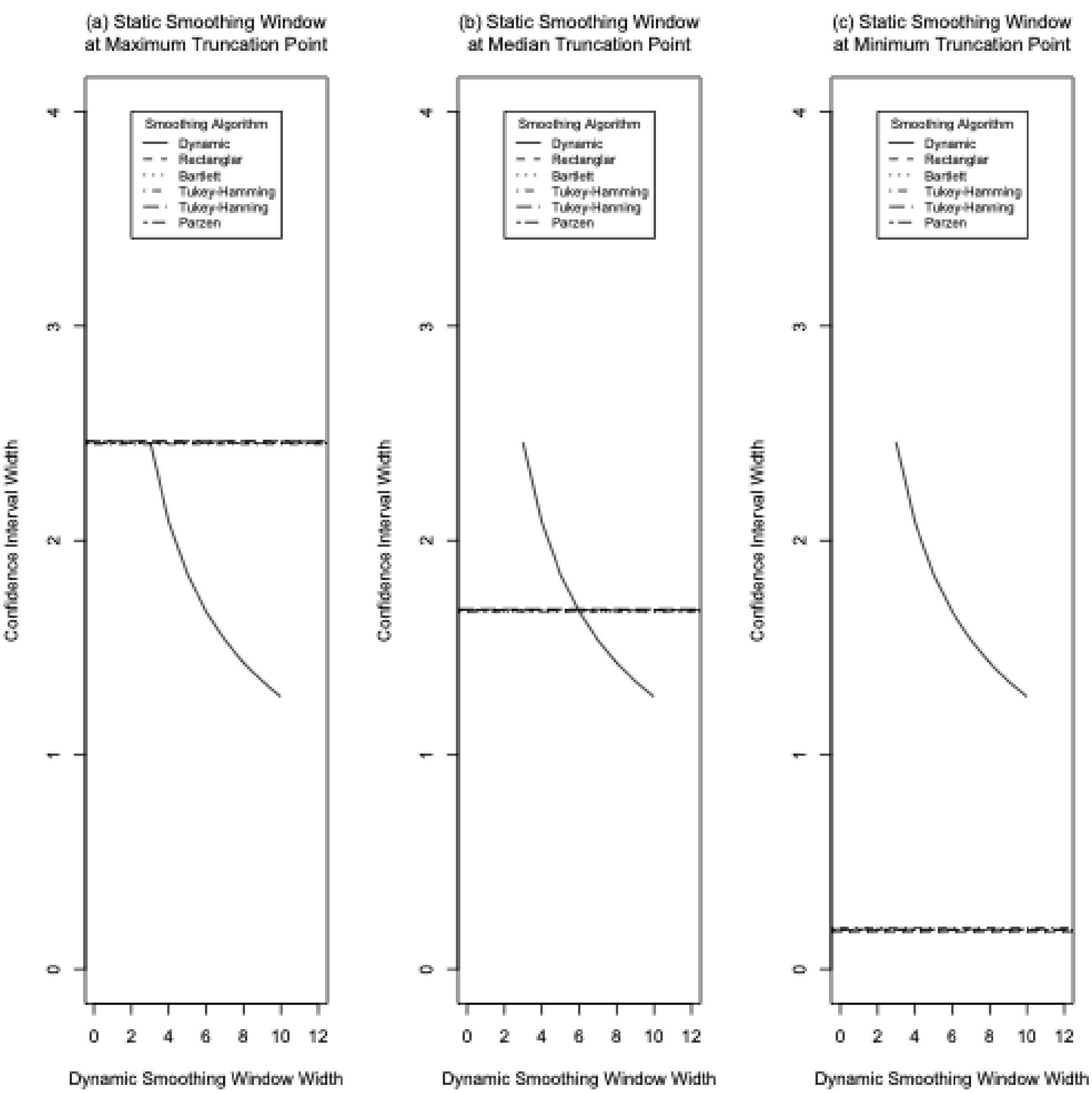

**FIGURE 2.3.4. A comparison of CI widths across possible dynamic smoothing window widths to CI widths for static smoothing window widths when $n = 1000$ and dynamic smoothing Proportion of Smoothing $= 0.01$.**

With respect to static smoothing, the confidence interval widths of the five static smoothing windows are close, but not identical, in size at each level of the truncation point (i.e., maximum, median, minimum). This occurs because the estimated degrees of freedom for each static smoothing window are, likewise, close, but not identical. It is also evident that as the truncation points decrease from maximum to minimum, the confidence interval widths decrease, and, thus, precision of signal strength estimates increases.

**3. Discussion**

Although previous work has established the sensitivity, accuracy, resolution, and robustness of Kolmogorov-Zurbenko periodograms with dynamic smoothing in estimating signal frequencies, a static smoothing window and its smoothing window width can be selected such that its estimate of signal strength is more precise than that of a dynamic smoothing window, in the presence of a signal. Because the precision with which periodograms estimate signal strength is reflected in the width of their confidence intervals, we presented their statistical basis, then developed candidate functions to compute and plot confidence intervals for the Kolmogorov-Zurbenko periodogram with dynamic smoothing. Examples of their use were provided for two scenarios: one with a single signal embedded in a high level of random noise and another with two signals – one relatively strong and one relatively weak – close in frequency and embedded in a high level of random noise. As expected, confidence interval widths were wide in the presence of a signal and narrow when no signal was present.

We then established and compared the theoretical and practical limits for the precision of signal strength estimates for Kolmogorov-Zurbenko periodograms with dynamic smoothing to standard log-periodograms with static smoothing by contrasting their respective confidence interval widths. In general, the confidence interval width for a log-periodogram is:

$$CI\ Width = ln\left(\frac{\chi^2_{\alpha/2}(\nu)}{\chi^2_{1-\alpha/2}(\nu)}\right)$$

From this Equation, it was clear that the greater the degrees of freedom, $\nu$, for a given smoothing window, the smaller the confidence interval width and, thus, the more precise the estimate of signal strength.

Of particular interest is the precision of signal strength estimate in the presence of a signal. For Kolmogorov-Zurbenko periodograms with dynamic smoothing, we showed that the theoretical limit of window width in the presence of a signal is 3, with a spectral smoothing window truncation point of $m = 1$ and a theoretical lower limit for its degrees of freedom of $\nu = 6$. Consequently, any static smoothing window with degrees of freedom $\nu > 6$ will be more precise. Because degrees of freedom, $\nu$, is determined by the number of observations, $n$, and the static smoothing window's related truncation point, $M$, it was possible to identify the upper bound to the truncation point such that a given static smoothing window would have a smaller confidence interval and, thus, be more precise, than a dynamic smoothing window. In addition, a static smoothing window must also be able to resolve two frequencies as well as or better than a dynamic smoothing window and this led to the identification of a lower bound for its truncation point. Taken together, the lower bound and the upper bound result in a set of conditional truncation points for each static smoothing window under consideration such that their respective estimates of signal strength are more precise than those of dynamic smoothing windows, yet are able to resolve signals close in frequency as well as dynamic smoothing windows.

However, a caveat was also discovered. In order for a static smoothing window to estimate signal strength more precisely than a dynamic smoothing window, yet maintain the same level of resolution, the difference between the two closest frequencies must be greater than $\frac{6\pi}{n}$, where $n$ is the number of observations in the time series. If this is not met, the static smoothing window cannot resolve these frequencies and the analyst should use a Kolmogorov-Zurbenko periodogram with dynamic smoothing to both identify signal frequencies and estimate their respective signal strength.

Given the ability of log-periodograms with static smoothing to both estimate signal strength with greater precision, while maintaining a high level of resolution, one may well ask why Kolmogorov-Zurbenko periodograms should be used in the first place. In answer, it should be used because of its heightened sensitivity to signals afforded by the dynamic smoothing window. This sensitivity allows the analyst to detect signals that would otherwise be obscured by random noise. If one cannot detect a signal, then accuracy and resolution of signal frequencies as well as precision of signal strength estimates are of no use.

Given the keen sensitivity, accuracy, and resolution of Kolmogorov-Zurbenko periodograms with dynamic smoothing in detecting, identifying, and separating signal frequencies and given the greater precision of signal strength estimates and commensurate frequency resolution that can be obtained using log-periodograms with static smoothing under requisite conditions, spectral analysis should follow a two-step protocol. First, analysts should use the Kolmogorov-Zurbenko periodogram with dynamic smoothing to detect, identify, and separate signal frequencies and, second, should use a Kolmogorov-Zurbenko periodogram with static smoothing to estimate signal strength and to compute the corresponding confidence intervals at those identified frequencies. In this way, they can detect, identify, and separate signal frequencies and can be precise in estimating signal strength. By using such an approach, analysts can better model and predict future values of a given variable that has an underlying periodicity in time, in space, or in both time and space. To this end, the next version of the *kza* package in *R* should include not only our candidate functions for computing and plotting Kolmogorov-Zurbenko periodograms with dynamic smoothing that include confidence intervals (i.e., *smoothWithCIs.kzp*, *plotWithCIs.kzp*), but also an additional option for the method argument that carries out static smoothing with a choice of static windows, including their corresponding confidence intervals.

# APPENDIX 1: *smoothWithCIs.kzp( )* Function

```
1  smoothWithCIs.kzp<-function(object, log=TRUE, smooth_level=0.05,
2  method = "DZ", CIalpha = 0.05)
3  {
4        if (class(object)!='kzp') stop ("Object type needs to be kzp.")
5        n<-length(object$periodogram)
6        spg<-rep(0,n)
7        m<-rep(0,n)
8        M<-rep(0,n)
9        CIupper<-rep(0,n)
10       CIlower<-rep(0,n)
11
12       if (log==TRUE) p=log(object$periodogram) else p=object$periodogram
13       if (method == "DZ") q<-variation.kzp(p)
14       else if (method == "NZ") q<-nonlinearity.kzp(p)
15
16       cc<-smooth_level*q$total
17
18       for ( i in (1:n) ) {
19                m[i]<-sum(q$matrix[i,1:n]<=cc)
20                spg[i]<-mean(p[(max(1,(i-m[i]+1))):(min(n,(i+m[i]-1)))])
21                M[i]<-length(p[(max(1,(i-m[i]+1))):(min(n,(i+m[i]-1)))])
22                CIupper[i]<-spg[i] + log((2*M[i])/qchisq((CIalpha/2), 2*M[i]))
23                CIlower[i]<-spg[i] + log((2*M[i])/qchisq((1-(CIalpha/2)), 2*M[i]))
24       }
25
26       object$smooth_periodogram<-spg
27       object$smooth_periodogram_CI_upper<-CIupper
28       object$smooth_periodogram_CI_lower<-CIlower
29       object$smooth_method=method
30       return(object)
31       }
```

**NOTES: Changes to existing *smooth.kzp( )* function to create *smoothWithCIs.kzp( )* function**

Line 2: addition of *CIalpha*, the level of *alpha* set by the analyst, as an argument in function
Line 8: initialization of *M*, a vector for window widths across the spectrum
Line 9: initialization of *CIupper*, a vector for confidence interval upper limits across spectrum
Line 10: initialization of *CIlower*, a vector for confidence interval lower limits across spectrum
Line 21: extraction of vector window widths across spectrum
Line 22: computation of confidence interval upper limits across spectrum using Equation 3.2.1.8
Line 23: computation of confidence interval lower limits across spectrum using Equation 3.2.1.8
Line 27: inclusion of vector of confidence interval upper limit values in returned object
Line 28: inclusion of vector confidence interval lower limit values in returned object

## APPENDIX 2: *plotWithCIs.kzp( )* Function

```
1  plotWithCIs.kzp <- function(x, ...)
2  {
3        if (is.null(x$smooth_periodogram)) dz<-x$periodogram
4        else dz<-x$smooth_periodogram
5        dzU<-x$smooth_periodogram_CI_upper
6        dzL<-x$smooth_periodogram_CI_lower
7        omega<-(0:(length(x$periodogram)-1))/x$window
8        plot(omega, dz-mean(dz), type="l", xlab="Frequency", ylab="",
9        ylim=c(min(dzL-mean(dz))-1,max(dzU-mean(dz))+1))
10       lines(omega, dzU-mean(dz), type="l", col="blue")
11       lines(omega, dzL-mean(dz), type="l", col="red")
12 }
```

**NOTES: Changes to existing *plot.kzp( )* function to create *plotWithCIs.kzp( )* function**

Line 5: import of vector of confidence interval upper limits values from object generated by *smoothWithCIs.kzp( )* function

Line 6: import of vector of confidence interval lower limit values from object generated by *smoothWithCIs.kzp( )* function

Line 9: inclusion of lower and upper limits of y-axis based on minimum value for confidence interval lower limits and maximum value for confidence interval upper limits

Line 10: addition of upper confidence interval line to Kolmogorov-Zurbenko periodogram with DiRienzo-Zurbenko algorithm or Neagu-Zurbenko algorithm smoothing

Line 11: addition of lower confidence interval line to Kolmogorov-Zurbenko periodogram with DiRienzo-Zurbenko algorithm or Neagu-Zurbenko algorithm smoothing